\documentclass[12pt]{article}

\input epsf
\def\beq{\begin{eqnarray}}
\def\eeq{\end{eqnarray}}
\def\m{M_*}
\def\mpl{M_{\rm Pl}}

\def\lsim{\mathrel{\rlap{\lower3pt\hbox{\hskip0pt$\sim$}}
     \raise1pt\hbox{$<$}}}         %less than or approx. symbol
\def\gsim{\mathrel{\rlap{\lower4pt\hbox{\hskip1pt$\sim$}}
     \raise1pt\hbox{$>$}}}         %greater than or approx. symbol

\begin{document}

\begin{titlepage}

\begin{flushright}
{ NYU-TH-04/03/22}
\end{flushright}
\vskip 0.9cm

\centerline{\Large \bf  Weakly-coupled metastable graviton}

\vskip 0.7cm
\centerline{\large Gregory Gabadadze}
\vskip 0.3cm
\centerline{\em Center for Cosmology and Particle Physics}
\centerline{\em Department of Physics, New York University, New York, 
NY, 10003, USA}

\vskip 1.9cm

\begin{abstract}

A graviton of a nonzero mass and decay width propagates five 
physical polarizations.  The question of 
interactions of  these polarizations is crucial for viability 
of models of massive/metastable gravity. This question is addressed
in the context of the DGP model of a metastable graviton. 
First, I argue that the well-known breakdown of a naive perturbative 
expansion at a low scale is an artifact of the weak-field expansion  
itself.   Then, I propose a different expansion -- the 
constrained perturbation theory -- in which the breakdown does not 
occur and the theory is perturbatively tractable all the way up 
to its natural ultraviolet cutoff.
In this approach the couplings of the extra polarizations to  
matter and their selfcouplings  appear to be suppressed and 
should be neglected in measurements at sub-horizon scales. 
The model reproduces  results of General Relativity at observable 
distances with high accuracy, while  predicting deviations from them 
at the present-day horizon scale.

\end{abstract}

\vspace{3cm}

\end{titlepage}

\section{Introduction}

An idea that the observable acceleration of the Universe could be a 
result of large-distance modification of gravity 
is attractive, and is  experimentally 
testable \cite {DDG}--\cite{Nima}. Moreover, large-distance 
modifications of 
gravity give rise to a conceptually new approach to the long-standing 
cosmological constant problem  \cite {DGS,ADDG}. Hence, development 
of the models of modified gravity becomes an important task.  

The DGP model \cite {DGP} is a covariant theory 
of the large-distance modification of gravity 
(see, e.g., \cite {Dick}--\cite{Padila}). 
Interactions in this model 
are mediated by a single graviton that lives in infinite-volume 
five-dimensional space-time. This graviton resembles 
a massive 4D spin-2 state since it has five  polarizations.
Experimental constraints on extra 
light states with gravitational interactions are 
rather severe, therefore, the question of how 
these extra polarizations interact with 
observable matter and with themselves becomes 
crucial for the consistency of the  model.

From a 4D perspective, the graviton behaves as a state of a 
nonzero decay width\footnote{ In 6D or higher \cite {DG}, 
it behaves as a massive graviton with a decay width   
\cite {Wagner,Kiritsis,DHGS}.}. As such it 
shares certain properties of a massive graviton (and 
massive non-Abelian gauge fields). 
If non-Abelian gauge fields  are prescribed 
mass $m_V$  by hand (i.e., without using the Higgs mechanism), 
then the nonlinear amplitudes of the theory 
become strong at a scale $m_V/g$, where $g$ is a gauge-coupling 
constant.  A similar effect exists in  a theory of gravity in which 
mass $m_g$  is introduced by hand as in Ref. \cite {PF}.  
This was shown in Ref. \cite {Arkady},
by calculating nonlinear corrections to a spherically symmetric 
static body of the gravitational radius $r_g$, 
in which case the weak-field approximation breaks down 
at a scale $\Lambda_m\sim m_g/ (r_gm_g)^{1/5}$ \cite {Arkady}. 
This breakdown  is a universal property of the perturbative expansion 
in the massive theory as it can be understood by looking  
at tree-level trilinear graviton vertex diagrams \cite {DDGV}.
A simplest trilinear (graviton)$^3$ diagram gives rise to 
the $1/m_g^4$ singularity \cite {DDGV}. For small $m_g$ 
this diagram is enhanced even though it is 
multiplied by an extra power
of the Newton constant.   This can also be understood in terms of
interactions of the longitudinal polarizations 
becoming strong \cite {AGS}. For the pure gravitational sector 
itself  the corresponding scale $\Lambda_m$  reduces 
to $m_g/ (m_g/\mpl)^{1/5}$, which can be made only as 
big as $m_g/ (m_g/\mpl)^{1/3}$ by adding higher 
nonlinear terms \cite{AGS}. 

The reason for the breakdown of perturbation theory at a low scale 
can be traced back to terms in the graviton propagator 
that contain  products of the structure
\beq
{p_\mu \,p_\nu\over m_g^2} \,
\label{sing}
\eeq
with similar structures or with the flat space metric. 
These terms do not manifest themselves in 
physical amplitudes at the linear level since they are multiplied 
by conserved currents, however,  they enter nonlinear diagrams 
leading to the breakdown of perturbation theory  for 
massive non-Abelian gauge fields or massive gravity
\footnote{Unfortunately, the  massive gravity in 4D 
\cite {PF} is an unstable  theory 
\cite {Deser} with an instability 
time scale that can be rather short \cite {GG}.}.

In the model of Ref. \cite {DGP} the graviton decay width 
is introduced by adding to General Relativity 
a term that is reparametrization invariant (see the 
action (\ref {1})). As a result, the terms similar to (\ref {sing}) 
in the propagator are gauge dependent.
In a simple gauge adopted in Ref. \cite {DGP} 
they read as follows:
\beq
{p_\mu \,p_\nu\over m_c p} \,,
\label{singDGP}
\eeq
where $m_c\sim 10^{-33}~{\rm eV}$ is a counterpart of $m_g$ and  
$p\equiv \sqrt{p^2}$ is a root of the Euclidean momentum 
square. Because of the singularity in $m_c$ in 
(\ref {singDGP}), perturbation theory breaks down 
precociously \cite {DDGV}. However, this breakdown is 
an artifact of an ill-defined perturbative expansion -- 
the  known  exact solutions of the model have no trace of 
the breakdown scale (see Ref. \cite {DDGV}). 
This shows that if one sums up all the tree-level 
perturbative diagrams, then the breakdown scale should disappear. 
A more ambitious program is to asks for 
quantum consistency of the model as an effective theory with a cutoff.
Such a consistency can only be established within the perturbation 
theory. The latter is ill-defined above the scale 
$m_g/ (m_g/\mpl)^{1/3}$ \cite {Luty} (see also \cite {Rubakov}), 
because of the same nonlinear diagrams that make 
the weak-field expansion ill-defined already at the 
classical level \cite {DDGV}. However, it is not clear 
whether the new scale that emerges in perturbation theory
has any physical significance. The perturbative breakdown 
at the quantum level  could be an artifact of the 
technical method itself.  As was recently argued  by 
Dvali \cite {Dvali}, in a certain 
toy sigma model a similar breakdown  
takes place, however,  the full resumed solution of the 
theory is valid well above the naive breaking scale.

In the next section I will discuss in detail the 
reasons for the breakdown of perturbation theory in the DGP model.
Then, in section 3 I will show how one can {\it  define} 
a different perturbative procedure
such that the singular terms like (\ref {singDGP}) are eliminated 
from the propagator  and the UV behavior of the amplitudes is 
regular all the way up to the cutoff of the theory.
The effective one-graviton exchange amplitude 
between two sources with the stress-tensors $T_{\mu\nu}$ that 
I will calculate in section 3 takes the form:
\beq
{\cal A}\,
=\,{1\over p^2\,+\,m_c\,p}\,\left ( T_{\mu\nu}^2\,-\,
{1\over 2}\,T^2 \,{ p^2\,+\,2\,m_c\,p \over p^2\,+\,3\,m_c\,p}\right)\,.
\label{Adgp0}
\eeq
This amplitude interpolates between the 
four-dimensional behavior at $p\gg m_c$
and the five-dimensional behavior at $p\ll m_c$. It has no poles
on a physical Riemann sheet and satisfies requirements of 
unitarity, analyticity and causality. 

Notice that the naive perturbation theory in the DGP model 
has no strong coupling scale if the localized 5D Einstein-Hilbert term 
is used on the brane worldvolume \cite {GS}   
and in the ``dielectric'' regularization of the 
DGP model \cite {Massimo2}. In higher dimensional 
generalizations of the DGP model
\cite {DG,Massimo1}, the perturbation theory is well-defined and 
there is no problem in the first place as was recently shown in 
\cite {GS}. The concern of the present work is the five-dimensional
model with the four-dimensional induced term (see the action (\ref {1}) 
below).

\section{What's wrong with perturbative expansion?}

A brief answer to the above question is as follows.  
The DGP model  has two {\it dimensionful} parameters:
the Newton constant $G_N$  and the graviton lifetime 
$m_c$. As a result, the naive perturbative expansion in 
powers of $G_N$ is contaminated by powers of $1/m_c$.
Hence, for small values of $m_c$ perturbation theory 
breaks down for the unusually low value of the energy scale.
The question is whether this breakdown is an 
artifact of perturbation theory, or it could be 
that the breaking scale is truly a physical scale 
at which the model needs certain UV completion. 
Depending on a concrete model at hand, the either
of above two possibilities  could  be realized.

To study these issues in detail we consider the 
action of the DGP model \cite {DGP}
\beq
S\,=\,{\mpl^2 } \,\int\,d^4x\,\sqrt{g}\,R(g)\,+
{\m^{3} }\,\int \,d^4x\,dy\,\sqrt{{\bar g}}
\,{\cal R}_{5}({\bar g})\,,
\label{1}
\eeq
where $R$ and ${\cal R}_{5}$ are the four-dimensional
and five-dimensional Ricci scalars, respectively, and
$\m$ stands for the gravitational scale
of the bulk theory. The analog of the graviton mass
is $m_c= 2\m^3/\mpl^2$. The higher-dimensional and
four-dimensional  metric tensors are related as
\beq
{\bar g}(x,y=0)\equiv g(x)\,.
\label{bargg}
\eeq
There is a boundary (a brane) at $y=0$ and ${\bf Z}_2$
symmetry across the boundary is imposed. The presence of the 
boundary Gibbons-Hawking term is implied to warrant  
the correct Einstein equations in the bulk.
Matter fields are assumed to be localized on a 
brane and at low energies, that we observe, 
they do not escape into the bulk. Hence, the matter action 
is completely four-dimensional $S_M=\int d^4x L_M$.
Our conventions are: $ \eta_{AB} = 
{\rm diag}\, [+----]\,; A,B = 0,1,2,3,{\it 5}\,;\mu,\nu = 0,1,2,3.$

The simplest problem is to calculate the Green's function 
$D^{\mu\nu;\alpha\beta}$ and the amplitude of interaction 
of two sources $T_{\mu\nu}$ on the brane
\beq
{\cal A}_{\rm 1-graviton}\, \equiv 
\,T_{\mu\nu}\,D^{\mu\nu;\alpha\beta}\,
T_{\alpha \beta}\,.
\label{A}
\eeq
In order to perform perturbative calculations one has to fix a  
gauge. A simple way  adopted in \cite {DGP} is to use  
harmonic gauge in the bulk $\partial^A h_{AB}= \partial_B h^C_C/2$. 
In this case the one-graviton exchange amplitude on the brane 
(in the momentum space) takes the form:
\beq
{\cal A}_{\rm 1-graviton}(p,\,y=0)\,=\,{T^2_{1/3} \over p^2\,+\,m_c\,p}\,,
\label{A13}
\eeq
where we denote the Euclidean four-momentum square by $p^2$:
\beq
p^2\,\equiv\,-p^{\mu}p_\mu \,=\,-p_0^2\,+\,p_1^2\,+ p_2^2\,+p_3^2\,\equiv
\,p_4^2\,+\,p_1^2\,+ p_2^2\,+p_3^2\,,
\label{psquare}
\eeq
and
\beq
T^2_{1/3} \,\equiv\, 8\,\pi\,G_N \left ( 
T^2_{\mu\nu}\,-\,{1\over 3}\,T\cdot T \right )\,.
\label{T1third}
\eeq
We can see that ${\cal A}_{\rm 1-graviton}$ in (\ref {A13}) 
is non-singular in the $m_c\to 0$ limit (moreover, the pole in the 
amplitude is on a nonphysical Riemann sheet,
see discussions in the next section.). 
However, the gauge dependent part of the propagator 
$D^{\mu\nu;\alpha\beta}$ contains terms proportional to 
(\ref {singDGP}).  These terms do not enter the one-graviton 
exchange amplitude, but, they do contribute to higher-order  
tree-level non-linear diagrams which
blow up in the $m_c\to 0$ limit \cite {DDGV}.

The same fact is reflected in the expression for the trace of 
$h_{\mu\nu}\equiv g_{\mu\nu}-\eta_{\mu\nu}$ 
which in the harmonic gauge takes the form \cite {DGP}:
\beq
{\tilde h}^\mu_\mu(p, y=0) \,=\,-\,{T \over 3\,m_c\,p}\,.
\label{trace}
\eeq
(Hereafter the tilde-sign denotes Fourier-transformed quantities
and we put $8\,\pi\,G_N\,=1$.). From this expression we learn 
that: (i) ${\tilde h}^\mu_\mu $ 
is a propagating field in this gauge;
(ii) ${\tilde h}^\mu_\mu $ propagates as a 5D field, 
i.e., it does not see the brane kinetic term; (iii) The 
expression for ${\tilde h}^\mu_\mu $  is singular in the 
limit $m_c\to 0$.  The gauge dependent part of the 
momentum-space propagator ${\tilde D}(p,y)$ 
contains the terms $p_\mu \,p_\nu {\tilde h} $, which, due to 
(\ref {trace}), give rise to the 
singular term  (\ref {singDGP}). Hence, to understand the origin of the 
breakdown of perturbation theory, one should look at the 
origin of the $1/m_c$ scaling in (\ref {trace}).

The singular behavior of ${\tilde h}^\mu_\mu $  
is a direct consequence of the fact that the four-dimensional 
Ricci curvature $R(g)$  in the linearized approximation 
is forced to be zero  by the  $\{55\}$ and/or 
$\{\mu 5\}$ equations of motion. This can be seen by direct 
calculation of $R$ and of those equations, but  it is more instructive 
to see this by using the ADM decomposition.  The $\{55\}$ equation reads:
\beq
R\,=\,(K^\nu_\nu)^2\,-\,K_{\mu\nu}^2\,,
\label{ADM}
\eeq
where $K_{\mu\nu}$ denotes the extrinsic curvature. Since  
$K\sim {\cal O} (h)$ the above equation implies that 
the four-dimensional curvature $R\sim {\cal O} (h^2)$ and  in the 
linearized order $R$ vanishes.  Let us now see how this leads to 
the singular behavior of  $h$ in (\ref {trace}). The junction 
condition across the brane  contains  two types of terms: 
there are  terms proportional to $m_c$ and there are terms that 
are  independent of $m_c$. The former come from the bulk Einstein-Hilbert 
action while the latter appear due to the worldvolume Einstein-Hilbert 
term. In the trace of the junction condition
the $m_c$--independent term is  proportional to the four-dimensional 
Ricci scalar $R$. On the other hand, as we argued above, $R$ has no 
linear in $h$ term in the weak-field expansion, simply because 
these terms cancel out  due to the  $\{55\}$ and/or 
$\{\mu 5\}$ equations.  Therefore, in the linearized approximation 
the junction condition  contains only the terms that  
come from the bulk. These terms are proportional to 
$m_c h$. This inevitably leads to the trace of $h$ (\ref {trace}) 
that is singular in the $m_c\to 0$ limit and 
triggers the breakdown of the perturbative approach as discussed above.

The above arguments  suggest  that the two limiting procedures, 
first truncating the small $h$ expansion and only then taking  
the $m_c\to 0$ limit, do not commute with each other. 
Therefore, the right way to perform the calculations is either 
to look at exact solutions of 
classical equations of motion, as was argued in 
\cite {Arkady,DDGV}, or to retain at least 
quadratic terms in the equations. The obtained results 
won't be singular in the $m_c\to 0$ limit.

However, neither of the above approaches addresses the 
issue of  quantum gravitational loops. Since the loops can only be 
calculated within a well-defined perturbation theory, 
one needs to construct a new perturbative expansion 
that would make diagrams tractable at short distances.

In the next section we will propose  to rearrange 
perturbation theory in such a way that the consistent 
answers be obtained in the weak-field approximation.  
This can be achieved if the linearized gauge-fixing terms 
can play  the role similar to the nonlinear terms. 
We will see that this  requires a certain nontrivial   
procedure of gauge-fixing and choosing of appropriate 
boundary conditions.

\section {Constrained perturbative expansion}

Below we develop a perturbative approach
that allows to perform calculations in the weak-field 
approximation without breaking the expansion at a low scale.

We recall that in the DGP model the boundary (the brane) 
breaks explicitly  translational invariance in the 
$y$ direction, as well as the rotational symmetry that 
involves the $y$ coordinate.
However, this fact is not reflected in the linearized 
approximation -- the linearized theory that follow from 
(\ref {1}) is invariant under five-dimensional 
reparametrizations\footnote{If instead of the boundary 
we consider a dynamical brane of a nonzero tension,  
then the five-dimensional Poincare symmetry is nonlinearly 
realized and one has to include a Nambu-Goldstone mode on the brane.}.
This line of arguments suggests 
to introduce constraints in the linearized theory that would account
for the broken symmetries.  It is clear that an arbitrary set 
of such constraint cannot be consistent with equations of 
motion with boundary conditions on the brane and at $y\to \infty$. 
However, by trial and error a 
consistent set  of constraints and gauge conditions can be found.
Below we introduce this set of equations step by step.
We start by imposing the following condition:
\beq
B_\mu \, \equiv\, \partial_\mu h_{55} \,+\,
\partial^\alpha h_{\alpha \mu}\,
=\,0 \,.
\label{bmu}
\eeq
Furthermore,  to make the kinetic term for the 
$\{\mu 5\}$ component invertible we set a second condition:
\beq
B_5\,\equiv\,\partial^\mu\,h_{\mu 5}\,=\,0\,.
\label{b5}
\eeq 
At a first sight, the two conditions 
(\ref {bmu}) and  (\ref {b5}) fix all the $x$-dependent gauge 
transformations and make the gauge kinetic terms 
non-singular and invertible. However, at a closer inspection this 
does not appear to be satisfactory.  
One can look at the $\{\mu\nu \}$ component of the equations of 
motion  and integrate this equation w.r.t. $y$  
from $-\epsilon $ to $\epsilon$,
with $\epsilon \to 0$. After the integration,  all the terms 
with $B_\mu$ and $B_5$ vanish. The resulting equation (which is just 
the Israel junction condition) taken by its own, is 
invariant under the following four-dimensional transformations
\beq
h^\prime_{\mu\nu}(x^\prime, y)|_{y=0}\,=\,h_{\mu\nu}(x, y)|_{y=0}\,+\,
\partial_\mu \zeta_\nu |_{y=0}\,+\, 
\partial_\nu \zeta_\mu |_{y=0}\,. 
\label{branetransf}
\eeq
This suggests that in the $m_c\to 0$ limit 
the gauge kinetic term on the brane is not invertible.
As a result, the problem of a precocious breakdown of perturbation theory
discussed in the previous section arises. 
To avoid this difficulty one can  introduce the following term 
on the brane worldvolume:
\beq
\Delta S\, \equiv \, 
-\, \mpl^2\,\int\,d^4x \,dy \,\delta (y)\,\left ( \partial^\mu h_{\mu\nu} - 
{1\over 2} \partial_\nu h^{\alpha}_{\alpha} \right )^2\,.
\label{gauge}
\eeq
This makes the graviton kinetic term of the brane invertible even in the 
$m_c \to 0$ limit. At this stage, the partition function can 
be {\it defined} as follows: 
\beq
Z_{\rm gf}\,=\,{\rm lim}_{\alpha,\gamma\to 0}
\int dh_{AB} \, 
{\rm exp} \,i \left ( S + \Delta S\,+ \m^3\int d^4x dy 
\, \left \{{B_5^2\over 2\gamma} + {B_\mu^2 \over 2\alpha} 
\right \} \right )\,.
\label{zgf}
\eeq
Here $S$ and $\Delta S$  are given in (\ref {1}) and 
(\ref {gauge}) respectively, 
and the limit $\alpha,\gamma\to 0$ enforces 
(\ref {bmu}) and (\ref {b5}). 
Before proceeding  further, notice that 
Eqs. (\ref {bmu}) and (\ref {b5}) would have
been just gauge-fixing conditions 
if the boundary were absent (e.g., in a pure 5D theory with no brane). 
However, in the present case, the above equations,
when combined with the junction condition across the brane, 
enforce certain boundary conditions on the brane. Therefore,  
Eqs. (\ref {bmu}) and (\ref {b5}) do more than 
gauge-fixing, and  $\gamma$ and $\alpha $ 
cannot be regarded as gauge fixing parameters. 
The prescription given by (\ref {zgf}) is to 
calculate first all Green's functions and then take the limit 
$\alpha, \gamma \to 0$.  Because of this, the results of 
the present calculations differ from \cite {DGP} where 
other boundary conditions were implied.

Using (\ref {zgf}) we calculate below 
the propagator $D$ and the amplitude ${\cal A}$ defined in (\ref {A}).
We will see that there are no terms in $D$ that blow up as $m_c\to 0$.

We start with the equations of motion that follow from 
(\ref {zgf}). The $\{\mu\nu\}$ equation on the brane reads
\beq
&{m_c\over 2}& \int_{-\epsilon}^{+\epsilon}\,dy \,
\left ( \partial_D^2\,h_{\mu\nu}\,-\,
\eta_{\mu\nu}\,\partial_D^2\,h^\alpha_\alpha  \,
+\,\partial_\mu\partial_5 h_{5\nu}\,+\,
\partial_\nu\partial_5 h_{5\mu}\, - \,2\, \eta_{\mu\nu}
\partial^\alpha \partial_5 h_{5\alpha}\, \right )+ G_{\mu\nu}^{(4)}
\nonumber \\ &-&
 (\partial_\mu \partial_\alpha h_{\alpha \nu} 
+\partial_\nu \partial_\alpha h_{\alpha \mu} 
- \partial_\mu \partial_\nu h^\alpha_\alpha -\eta_{\mu\nu}
 \partial_\alpha \partial_\beta h^{\alpha \beta} +{1\over 2} 
 \eta_{\mu\nu} \partial_4^2 h^\alpha_\alpha  ) = T_{\mu\nu},
\label{munubrane}
\eeq 
where $\epsilon \to 0$,  
$\partial_D^2\equiv \partial_A\partial^A$,
$\partial_4^2\equiv \partial_\mu\partial^\mu$.
In (\ref {munubrane}) we retained only 
terms that are nonzero in the  $\epsilon \to 0$ limit.
Furthermore, $G_{\mu\nu}^{(4)}$ denotes the 4D Einstein tensor:
\beq
G^{(4)}_{\mu\nu}= \partial_4^2\,h_{\mu\nu}- \partial_\mu 
\partial_\alpha 
h^\alpha_\nu  - \partial_\nu  \partial_\alpha 
h^\alpha_\mu + \partial_\mu\,\partial_\nu\,h^\alpha_\alpha
%\,\nonumber \\[2mm]
-  \eta_{\mu\nu}
\partial_4^2\,h^\alpha_\alpha +  \eta_{\mu\nu}\,\partial_\alpha
\,\partial_\beta\,h^{\alpha \beta}.
\label{G4}
\eeq
The $\{\mu\nu\}$ equation in the bulk takes the form: 
\beq
& \partial_D^2\,h_{\mu\nu}& -\,\eta_{\mu\nu}\,\partial_D^2\,h^\alpha_\alpha
\,-\, \partial_\mu \partial^\alpha h_{\alpha \nu} \,-\,
\partial_\nu \partial^\alpha h_{\alpha \mu} \,+\,
\partial_\mu \partial_\nu h^\alpha_\alpha \nonumber \\ 
&+& \eta_{\mu\nu}\,\partial_\alpha \partial_\beta h^{\alpha \beta}\,
+\,\eta_{\mu\nu}\, \partial_4^2\,h_{55}\,-\,\partial_\mu\partial_\nu \,h_{55}
\,+\,\partial_\mu\partial_5 h_{5\nu}\nonumber \\ 
&+& \partial_\nu\partial_5 h_{5\mu}\, -\,2\, \eta_{\mu\nu}
\partial^\alpha \partial_5 h_{5\alpha}\,-\,{1\over \alpha}\,
( \partial_\mu\partial_\nu h_{55}\,+\, \partial_\mu \partial^\alpha
h_{\alpha \nu})\,=\, 0\,.
\label{munubulk}
\eeq
As a next step we turn to the $\{\mu 5 \}$ equation  which reads as follows: 
\beq
\partial_4^2 h_{\mu 5} \,-\,\partial_\mu \partial^\alpha h_{\alpha 5}
-\partial_5 (\partial^\alpha h_{\alpha \mu} -\partial_\mu h^\alpha_\alpha)
 - {1\over \gamma}\, (\partial_\mu \partial^\alpha h_{\alpha 5})\,=\,0\,.
\label{mu5}
\eeq
Finally, the $\{5 5 \}$ equation takes the form
\beq
\partial^2_4\, h^\alpha_\alpha -\partial_\mu\partial_\nu h^{\mu\nu}
-{1\over \alpha} (\partial^2_4\,h_{55} + \partial_\mu\partial_\nu h^{\mu\nu})
\,=\,0\,.
\label{55}
\eeq
After the calculation is done the limit
$\alpha, \gamma \to 0$ should be taken.
We turn to the momentum space w.r.t. four worldvolume coordinates:
\beq
{h}_{AB}(x, y)\, = \,\int d^4p \,e^{ipx}\, {\tilde h}_{AB}(p, y)\,.
\label{mom}
\eeq
From the above equations we calculate the response of gravity to the source 
$T_{\mu\nu}$. In the  limit $\alpha, \gamma \to 0$ the results are as follow:
\beq
{\tilde h}_{\mu\nu}(p, y)\, \to \, 
{1\over p^2\,+\,m_c\,p}\,\left ( T_{\mu\nu}\,-\,
{1\over 2}\,\eta_{\mu\nu}\,T \,{ p^2\,+\,2\,m_c\,p \over p^2\,+\,
3\,m_c\,p}\right)e^{-p|y|}\,.
\label{hmunu}
\eeq
We note that in this expression there are no terms similar 
to (\ref {singDGP}),  unlike to what happens in the 
harmonic gauge \cite {DGP} where the singular terms are present.

For the off-diagonal components we find that 
${\tilde h}_{\alpha 5}\sim \gamma \,p_\alpha $, and 
\beq
{\tilde h}_{\alpha5}(p, y) \to 0\,.
\label{hmu5}
\eeq
Finally,   
\beq
{\tilde h}_{55} \,\to - \,{r\over 2}
{\tilde h}^\alpha_\alpha \,\to  \,
{r\over 2}\,{T\over p^2+3m_cp}\,e^{-p|y|}\,,
\label{h55}
\eeq
with $r\equiv (p^2+2m_c p)/(p^2+m_cp)$.
The amplitude on the brane, as was already stated in 
(\ref {Adgp0}), takes the form
\beq
{\cal A}_{\rm 1-graviton}(p, y=0)\,
=\,{1\over p^2\,+\,m_c\,p}\,\left ( T_{\mu\nu}^2\,-\,
{1\over 2}\,T^2 \,{ p^2\,+\,2\,m_c\,p \over p^2\,+\,3\,m_c\,p}\right)\,.
\label{Adgp}
\eeq
A remarkable property of this amplitude is that 
it interpolates between the 4D behavior at $p\gg m_c$
\beq
{\cal A}_{4D}(p, y=0)\,
\simeq\,{1\over p^2 }\,\left ( T_{\mu\nu}^2\,-\,
{1\over 2}\,T^2 \right)\,,
\label{A4D}
\eeq
and the 5D amplitude at $p\ll m_c$
\beq
{\cal A}_{5D}(p, y=0)\,
\simeq \,{1\over m_c\,p }\,\left ( T_{\mu\nu}^2\,-\,
{1\over 3}\,T^2 \right)\,.
\label{A5D}
\eeq
This amplitude has no vDVZ  discontinuity \cite {Iwa,vdv,Zakharov}.

It is instructive to rewrite the amplitude (\ref {Adgp})  
in the following form:
\beq
{\cal A}_{\rm 1-graviton} \,=\,{T^2_{1/2} \over p^2\,+\,m_c\,p}\,
+\,{1\over 6}\,T^2 \,{g(p^2)\over p^2\,+\,m_c\,p}\,,
\label{Adgpsplit}
\eeq
where
\beq
T^2_{1/2} \,\equiv\,\left 
( T^2_{\mu\nu}\,-\,{1\over 2}\,T\cdot T \right )\,,
\label{T1half}
\eeq
and 
\beq
g(p^2)\,\equiv \, {3\,m_c\,p \over p^2\,+\,3\,m_c\,p}\,.
\label{gp}
\eeq
The first term on the r.h.s. of (\ref {Adgpsplit}) is 
due to  two transverse polarizations of the graviton, 
while the second term 
is due to an extra scalar polarization. The scalar
acquires a momentum-dependent form-factor. The form-factor is such that 
at sub-horizon  distances, i.e., when $p\gg m_c$,  
the scalar decouples. At these scales
the effects  of the extra polarization is suppressed by a factor 
$m_c/p$ (e.g., in the Solar system this is less than $10^{-13}$).
However, the scalar polarization 
kicks in at  super-horizon scales, $p\ll m_c$,  
where the five dimensional laws or gravity are restored. 
\begin{figure}
\centerline{\epsfbox{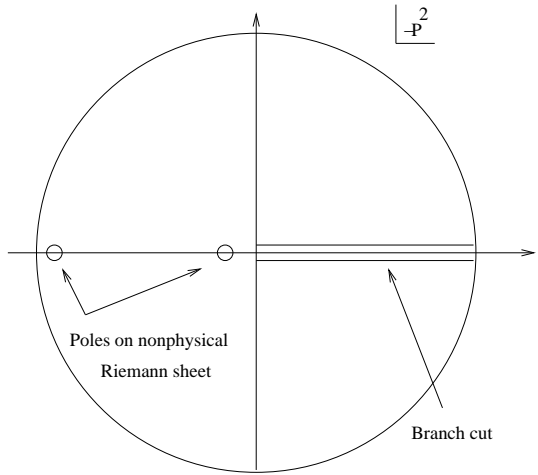}}
\epsfysize=6cm
\vspace{0.1in}
\caption{\small The complex plane of Minkowskian momentum square 
with the branch cut and poles on a second Riemann sheet. The  
physical Riemann sheet is pole free.}
\label{fig}
\end{figure} 

Let us discuss the above results in more detail. 
For this we study the pole structure of the amplitude  
(\ref {Adgpsplit}). There are two nontrivial poles  
\beq
p^2\,=\,-\,m_c\,p\,,~~~~{\rm and}~~~p^2\,=\,-3\,m_c\,p\,.
\label{poles}
\eeq
Let us find the positions of these poles 
on a complex plane of the Minkowskian momentum square $p_\mu^2$,
where $p^2=p_\mu^2 {\rm exp}{(-i\pi)}$. For this we note that 
there is a branch cut from zero to plus infinity on the 
complex plane (see Fig. 1). The pole at $p^2=0$ 
is just the origin of the branch cut. Because of the  cut the 
complex plane has many sheets (the propagator is multivalued 
function due to the square root in it). It is straightforward to show that 
both of the poles in (\ref {poles}) are on a {\it non-physical},
second Riemann sheet. Moreover, the positions of these poles
are far away from the branch cut (usual particle physics resonances
appear on non-physical sheets  close to the branch cut, 
the above poles, however,  are located on a negative semi-axis 
of the second Riemann sheet). Hence, the physical Riemann sheet is pole 
free\footnote{Because of the $\sqrt{p^2}$ dependence of the propagator
there are two choices of the sign of the square root. We choose
the sign as above.   For this choice the  poles do not appear
on the physical sheet and Euclidean Green's functions decay for 
large $y$. However, an opposite choice of the sign  of $\sqrt{p^2}$
can also be adopted. This would correspond to a 
different branch of the theory. On that branch, if we insist on flat 
brane, we find  tachyonic poles on a physical Riemann sheet.
This indicates that Minkowski space on that branch is unstable.
The unstable classical solutions found in Ref. \cite {Luty} do 
precisely correspond to this choice of the sign of the square 
root. On that branch one can also obtain the selfaccelerated  
solution without introducing the cosmological constant \cite {D,DDG}.  
This branch is decoupled (at least classically) from the branch that 
we are discussing in this work.}. The poles on a nonphysical sheet 
correspond to metastable states that do not appear as {\it in} and {\it out}
states in the S-matrix \cite {Veltman}. Using the contour of Fig. 1 
that encloses the plane with no poles, and taking 
into account the jump across the cut, the four-dimensional 
K\"allen-Lehmann  representation can be written for the 
amplitude (\ref {Adgpsplit}). The latter warrants four-dimensional 
analyticity, causality and  unitarity of the amplitude 
(\ref {Adgp})\footnote{{\it A priori} it is not clear why 
the theory that is truly higher-dimensional at all 
scales should have respected 4D analyticity and causality.}.
Although the above interpretation is the only correct one,
one could certainly adopt the following provisional picture 
that might be convenient for intuitive thinking. The second pole 
in (\ref {poles}) can be interpreted as a ``metastable ghost'' 
with a momentum-dependent decay width that accompanies the fifth 
polarization and cancels its contributions at short distances. 
Remarkably, this state does 
not give rise to  the usual instabilities because it  can only appear in
{\it intermediate states} in Feynman diagrams, 
but  does not appear in  the {\it in} and {\it out} states in the S-matrix 
elements. In this respect, it is more appropriate to think 
that the scalar graviton  polarization acquires the 
form-factor $g(p)$ (\ref {gp}).

The above results seem somewhat puzzling from the point of view of the 
Kaluza-Klein (KK) decomposition. Conventional intuition would suggest that 
the spectrum of the KK modes consists of massive spin-2 states. 
The K\"allen-Lehmann representation for the 
amplitude as a sum w.r.t. these  massive states 
would give rise to the tensorial structure  
where  the first term on the r.h.s. of 
(\ref {Adgpsplit}) is proportional to $T^2_{1/3}$, 
instead of $T^2_{1/2}$. 
In this case, the remaining part of the amplitude on the r.h.s. would have 
a {\it negative} sign. This might be thought of 
as a problem. However, this is not so. The crucial difference 
of the present approach from the conventional KK theories is that 
the effective 4D states are mixed states of an infinite number 
of tensor  and scalar modes. What is responsible for the mixing between  
the different spin states is the brane-induced term and 
the present procedure of imposing the constraints. 
In the covariant gauge that we discuss the 
trace of $h$ propagates and mixes with tensor fields.
From the KK point of view this would look as an infinite 
tower of states with wrong kinetic terms. However, at 
least in the linearized  approximation, the trace is a gauge  artifact 
(similar to the zeroth component of the gauge field in covariantly 
gauge-fixed QED or QCD).  Nevertheless, the effect of the trace part is 
that the true physical eigenmodes do not carry a definite 
four-dimensional spin of a local four-dimensional theory 
(see also \cite {Massimo2}).
Because of this  there is no reason to split the amplitude 
(\ref {Adgpsplit}) into the term that is proportional to $T^2_{1/3}$ 
and the rest.

The question of interactions of these states 
in the full  nonlinear theory is not addressed in the present work.  
What happens with the diagrams in which the  ``metastable ghosts'' 
propagate in the  loops (the unitarity cuts of which 
should give production of these multiple states) 
remains unknown. However, since the theory possesses 
4D reparametrization invariance, we expect that these questions will find 
answers similar to those of  non-Abelian gauge fields. Further  
studies are being conducted to understand these issues.

\section{Conclusions}

To summarize briefly, a new, 
{\it constrained perturbative expansion} was proposed.
In this approach perturbation theory is well-formulated.
The resulting amplitude interpolates between the 4D behavior at 
observable distances and 5D behavior at super-horizon scales.
This is due to the scalar polarization of the graviton 
that acquires a momentum-dependent form-factor. As a result, 
the scalar decouples with high accuracy from the 
observables at sub-horizon distances. 

The model can potentially evade the no-go theorem for massive/metastable  
gravity \cite {vdv}, that states that for the cancellation of the 
extra scalar polarization one should introduce a ghost that would give 
rise to instabilities \cite {vdv, DGP1,DGP2}. In the present model, at least
in the linearized approximation, such instabilities do not occur.
The convenient (although not precise)  
picture is to think of a ``metastable ghost'' that 
exists only as an intermediate state in Feynman diagrams 
which does not appear in the final states at least in the linearized
theory. Since this state cannot be emitted in physical processes, it 
does not give rise to the usual instability. The latter property 
is similar to the observation made in the ``dielectric'' regularization of the 
DGP model in \cite {Massimo2}. 

The questions that remain open concern the gauge-fixing and interactions 
in the full non-linear theory where the Faddeev-Popov 
ghosts are expected to play a crucial role. These issues will be 
addressed elsewhere. 

\vspace{0.3in}

I would like to thank Gia Dvali and Massimo Porrati 
for valuable comments, and Max Libanov for useful 
critical remarks.

\end{document}